\documentclass[12pt]{elsart}
 \usepackage{amsfonts}
 \usepackage{amssymb,amsxtra}
 \usepackage{amsmath}  
 \usepackage{graphicx,psfrag,epsfig,here}
 \usepackage{cite}

\newcommand{\Mpi}{M_{\pi}}

\newcommand{\beq}{\begin{equation}}
\newcommand{\eeq}{\end{equation}}
\newcommand{\bea}{\begin{eqnarray}}
\newcommand{\eea}{\end{eqnarray}}
\newcommand{\bdm}{\begin{displaymath}}
\newcommand{\edm}{\end{displaymath}}

\newcommand{\mr}{\mathrm}

\newcommand{\MeV}{\,\mr{MeV}}
\newcommand{\GeV}{\,\mr{GeV}}

\newcommand{\lr}[1]{\ell^{\, {\mr r}}_{#1}}
\newcommand{\hr}[1]{h^{{\mr r}}_{#1}}

\newcommand{\nn}{\nonumber\\}
\newcommand{\nnnl}{\nonumber\\}

\newcommand{\ep}{\epsilon}

\newcommand{\LN}[1]{\ln \left(#1 \right)}

\newcommand{\gla}{\langle}
\newcommand{\gra}{\rangle}

\newcommand{\ged}{\end{document}}

\renewcommand{\thesection}{\arabic{section}}
\renewcommand{\thesubsection}{\arabic{section}.\arabic{subsection}}
\renewcommand{\theequation}{\arabic{equation}}

\begin{document}

\begin{frontmatter}

\title{\Large\bf Chiral logarithms to five loops}

\author{M.~Bissegger$^1$ \and A.~Fuhrer$^2$}

\address{Institute for Theoretical Physics, University of Bern,
Sidlerstr. 5, CH-3012 Bern, Switzerland}

\thanks[bissegger]{bissegg@itp.unibe.ch}
\thanks[fuhrer]{afuhrer@itp.unibe.ch}

\begin{abstract}
We investigate two specific Green functions in the framework of chiral
perturbation theory. We  show that, using analyticity and unitarity,
their leading logarithmic singularities can be evaluated in the chiral limit to any desired order
in the chiral expansion, with a modest calculational cost. 
The claim is illustrated with an evaluation of the leading logarithm 
for the scalar two--point function to five--loop order.

\end{abstract}


\end{frontmatter}

\section{Chiral logarithms in the chiral limit}
In this article, we discuss the structure of chiral 
logarithms in the effective low--energy theory of QCD, chiral perturbation
theory \cite{Weinberg,GL1983}.
 To simplify the discussion, we consider the case of two 
flavours $u$ and $d$. The relevant effective 
Lagrangian is explicitly known  at next-to-next-to-leading order 
in the chiral expansion \cite{GL1983,BCE_1_1997,BCE_2_1997}. Its
 structure is
\beq
{\cal L} = {\cal L}_2+{\cal L}_4+{\cal L}_6+\ldots\,\,,
\eeq
where the indices stand for the chiral order of the Lagrangian.
We  count ${\cal L}_{n}$ 
a quantity of order $p^{n}$. The low--energy expansion amounts to an expansion
in powers of $p^2$. The terms of order $p^4$ and $p^6$ e.g. are  generated by
\beq\label{eq:powershbar}
\begin{array}{cccc}
&{\cal L}_2\times {\cal L}_2,&{\cal L}_4:&p^4,\\[2mm]
{\cal L}_2\times {\cal L}_2\times {\cal L}_2,&{\cal L}_2\times {\cal
  L}_4,&{\cal L}_6:&p^6,
\end{array}
\eeq
where the symbol $\times$ denotes a loop integral, and where
${\cal L}_2\times {\cal L}_2\times {\cal L}_2$ 
denotes a two--loop integral, with 3 vertices from ${\cal L}_2$, etc.
We furthermore use the convention that the  low--energy constants (LECs) in ${\cal L}_n, n\geq 4$ 
are made dimensionless by multiplying
 them with appropriate powers of $F^{-2}$, where $F$ denotes the pion decay
 constant in the chiral limit.
As usual, we call the terms of order $p^{2n+2}$ generically 
 $n$--loop contributions, although
they are not exclusively generated by  $n$ independent loop
integrals.

In the following, we consider the two--point function
of two scalar quark currents in the chiral limit $m_u=m_d=0$,
\beq\label{eq:scorrchi}
H(s)=
i\int dx e^{ipx}\langle 0|T S^0(x)S^0(0)|0 \rangle \,;\, S^0=\bar uu+\bar dd\,;\, s=p^2,
\eeq
whose low--energy expansion is of the form
\beq
H(s)=H_2+ H_4+H_6+\ldots\,,
\eeq
where $H_{n}$ is considered to be of  order $p^n$ according to
 the above counting.
There is no tree graph contribution in this case, $H_2=0$. 
The leading term is
\cite{GL1983}
\begin{align}
H_4&=B^2[A_1L+A_0]\,;\,L=\ln{-\frac{s}{\mu^2}}\,\,,\nnnl   
A_1&=-\frac{3}{8\pi^2}\,,\,A_0=\frac{3}{8\pi^2}+8(\lr{3}+\hr{1})\,.
\end{align}
The quantity $B$ is related to the quark condensate,
$\langle 0|S^0(0)|0\rangle=-2BF^2$, and $\lr{3},\hr{1}$ are renormalized 
LECs from ${\cal L}_4$. The symbol $L$ denotes a ${\it chiral\, logarithm}$
which is generated by the relevant one--loop graph, and $\mu$ stands for the
scale which is introduced using $d$--dimensional regularization of the 
loop integral. The scale dependence of 
$\lr{3}, \hr{1}$ cancel the scale dependence of the chiral logarithm -- the
quantities $H_n$ are scale independent.

\begin{figure}
\centering
\includegraphics[width=5cm]{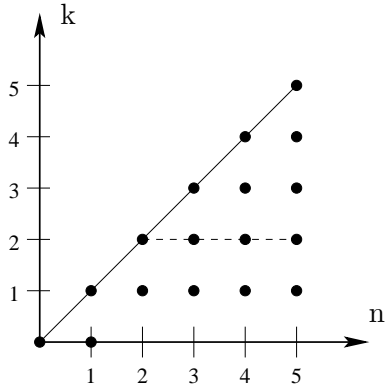}
\caption{The power $k$ of chiral logarithms of the correlator as a function of
  the number $n$ of loops. The solid line corresponds to leading logarithms and the
dashed line connects the terms of order $L^2$.\label{fig:nloops}}
\end{figure}

Higher orders in the low--energy 
expansion generate additional powers of $L$. The two--loop contribution
e.g. contains a term proportional to $L^2$,
\beq
H_6=\frac{sB^2}{F^2}[B_2L^2+B_1L+B_0]\,,
\eeq
where
\beq
B_2=\frac{3}{128\pi^4},
\eeq
see below. In addition to the square $L^2$, the two--loop term $H_6$ contains 
as well a single logarithm,
as the one--loop contribution $H_4$ does. However, this term is suppressed 
here by one power in $s$, and becomes 
negligibly small at small momenta with respect to the one in $H_4$.
 Likewise, the three--loop contribution $H_8$
contains $L^n, n\leq 3$, where 
the term proportional to $L^2$ is
again suppressed by one power of $s$ with respect to the one in $H_6$, etc.
The general structure of the loop expansion is illustrated in Fig. 
\ref{fig:nloops}, where we display the power of chiral logarithms 
as a function of the number of loops. 

In the following, we 
call the logarithm $L^n$ generated by
$n$--loop graphs {\it the leading logarithms} (LL). They come with power
$s^{(n-1)}$ in the chiral expansion of $H$ and correspond to the 
contributions connected by the tilted, solid line in 
Fig.~\ref{fig:nloops}. Terms of order $L^n$ generated by 
graphs with more than $n$ loops 
are   suppressed by additional powers
of $s$, see the horizontal, dashed line in the figure, 
which connects the terms of order $L^2$.

Performing a renormalization group analysis, it can  be shown
\cite{colangelobuechler} that the leading logarithm $L^n$ is generated 
by $n$--loop integrals of the type
 ${\cal L}_2\times{\cal L}_2\times \ldots \times {\cal L}_2$ ($n+1$ vertices from
 ${\cal L}_2)$. Its coefficient  does not, 
therefore, contain any LECs -- it is
 determined by ${\cal L}_2$ alone.

There are two obvious questions: Is it possible 
\begin{enumerate}
\item[i)]
to calculate the leading
logarithm for any $n$? 
\item[ii)] 
to sum up the leading logarithms, similarly to summing up 
leading logarithmic singularities in renormalizable theories?
\end{enumerate}

Here, we concentrate on the first question, the second question will be
addressed in a subsequent publication \cite{BF}. There are several methods to
evaluate the leading logarithms:
\begin{enumerate}
\item[i)]
 Calculation by brute force. This procedure can obviously
not be implemented beyond the first few terms. 
\item[ii)]
The renormalization group can be used to show that the leading 
logarithm $L^{n}$ can be determined by a one--loop calculation with the
Lagrangian ${\cal L}_{2n}$  for any $n\geq 2$
\cite{colangelobuechler}.
While this sounds promising, the
method cannot, in practice, be used to perform the calculation beyond the
three--loop level.
\item[iii)]
The third method makes use of the fact that the correlator 
$H(s)$ has a rather simple structure in the chiral limit. Unitarity,
 analyticity and the Roy equations then allow one to to determine 
the leading logarithms rather easily. 
\end{enumerate}
We now illustrate methods ii) and iii) in turn, and start with ii).

\section{Leading logarithms from a one--loop calculation}\label{sect:oneloop}
We illustrate the method with $H_8$. We make use of the fact
that $H_n$ are scale independent,
\beq\label{eq:scaleHn}
\dot H_n=0\,,
\eeq
where we define the dot--operation 
by $\dot X\doteq\mu\frac{dX}{d\mu}$.  
After renormalization, in four dimensions, $H_8$  has the
structure 
\beq
\renewcommand{\arraystretch}{1.5}
\begin{array}{llclclcl|l}
H_8= B^2 \bigg[&C_3 L^3&+&C_2L^2 &+&C_1 L& +
 &C_0& \quad \mathcal{L}_2 \times\mathcal{L}_2\times\mathcal{L}_2\times\mathcal{L}_2\\
&&+&D_2(\lr{i})L^2 &+&D_1(\lr{i})L &+&D_0(\lr{i})& \quad\mathcal{L}_2 \times \mathcal{L}_2
\times \mathcal{L}_4\\
&&&&+&E_1(\lr{i} \lr{k}) L& +&E_0(\lr{i} \lr{k})& \quad \mathcal{L}_4\times
\mathcal{L}_4\\
&&&&+&F_1(c^{r}_i) L &+&F_0(c^{r}_i)  & \quad \mathcal{L}_2 \times \mathcal{L}_6\\
&&&&&&+&G_0(d^r_i) \bigg]. & \quad \mathcal{L}_8 
\end{array}
\eeq
Here we have split the various contributions 
 into different components which
show the explicit dependence on the scale dependent LECs. According to the power counting
rules it is possible to assign these components to the corresponding type of
diagram. The polynomials $E_{ \{1,0 \} }(\ell_i \ell_k)$ for example stem from one loop diagrams
with two vertices from $\mathcal{L}_4$.
The $c_i^r$ denote renormalized LECs from ${\cal L}_6$.
 Scale independence demands 
\begin{align}\label{eq:log3}
C_3 &= \frac{1}{6} \dot{D}_2(\lr{i}),\nn
\dot{D}_2(\lr{i}) &= \frac{1}{4}\left(\ddot{E}_1(\lr{i} \lr{k}) +
  \ddot{F}_1(c^r_i)+\ddot{D}_1(\lr{i})  -4 \dot{C}_2 \right).
\end{align}
\begin{figure}
\centering
\begin{tabular}{cc}
\includegraphics[height=1.3cm]{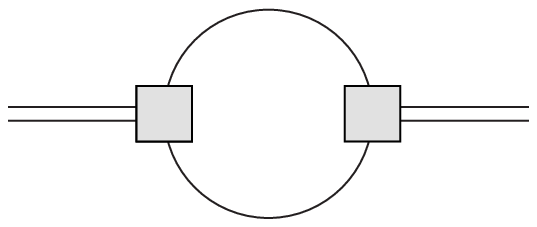}&\includegraphics[height=1.3cm]{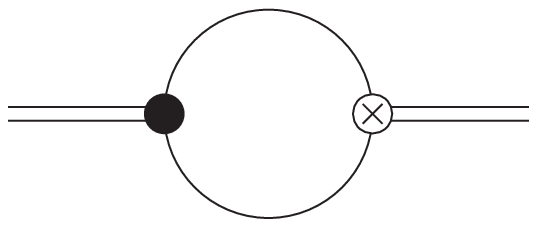}\\
$G_{\mathcal{L}_4\times\mathcal{L}_4}$&$G_{\mathcal{L}_6 \times \mathcal{L}_2}$
\end{tabular}
\caption{Diagrams used in the calculation of the three--loop leading
  logarithm of the scalar two--point function. The cross denotes a vertex from
  $\mathcal{L}_2$, the square denotes a vertex from $\mathcal{L}_4$ and the black
  dot stands for a vertex from $\mathcal{L}_6$. \label{fig:3loop}}
\end{figure}
In fact the last two terms vanish. 
$C_3$ is determined by $E_1(\lr{i} \lr{k})$ and $F_1(c^r_i)$ which are given
by one 
loop calculations with either two vertices from $\mathcal{L}_4$ or
one vertex from $\mathcal{L}_6$ and one vertex from $\mathcal{L}_2$.
Performing this calculation explicitly, one has to calculate the diagrams
indicated in Fig.~\ref{fig:3loop}. The diagram $G_{\mathcal{L}_4\times\mathcal{L}_4}$ yields the contribution 
\beq
G_{\mathcal{L}_4\times\mathcal{L}_4}(s) =  -6i B^2 \frac{s^2}{F^4} \ell_4^2 J(s)
\eeq
and for $G_{\mathcal{L}_6\times \mathcal{L}_2}$ one obtains
\beq
G_{\mathcal{L}_6 \times \mathcal{L}_2}(s) =  96i B^2 \frac{s^2}{F^2} c_6 J(s),
\eeq
with 
\beq
J(s) = \frac{\mu^{d-4}}{16 \pi^2}\Big[\frac{2}{4-d}+\Gamma'(1)-\LN{-\frac{s}{4 \pi
    \mu^2}}+2\Big].
\eeq
Considering the coefficient of the logarithm $L$ and using Eq.~(\ref{eq:log3}) one
finds for the leading logarithm to three loops
\beq\label{eq:A3}
C_3  = -\frac{61}{9}\frac{s^2}{(16 \pi^2)^3 F^4}+O(s^3).
\eeq
{\underline{Remark:}} 
The above argument
shows that, to calculate $C_3$, it is sufficient to 
know the Lagrangian ${\cal  L}_6$, whereas ${\cal L}_8$ 
would be needed to renormalize all three--loop diagrams. Indeed, 
using the same arguments as above, it can be shown that a tree--level
calculation with ${\cal L}_8$ suffices to calculate $C_3$. 
Likewise, a two--loop calculation with ${\cal L}_4$ suffices as well to
determine $C_3$. {\underline{End of remark}}.

Whereas this method to calculate the leading logarithms is simple in
principle, it soon becomes prohibitively complicated: To evaluate the term
$L^4$ by a one--loop calculation, one would need the renormalized ${\cal
  L}_8$. We now present a method that allows one to calculate the LLs to
arbitrary orders.

\section{Leading logarithms from unitarity and analyticity}
Here, it is useful to
slightly rearrange the loop expansion of the correlator.
We collect terms with the same power of the chiral logarithm and write
\beq\label{eq:Hpol}
H(s) = \frac{B^2}{16 \pi^2}\left\{P_0+P_1L
+P_2L^2+\cdots\right\},
\eeq
where $P_i$ denote (dimensionless)  polynomials in the variable $N = s/(16 \pi^2 F^2)$, and
in the scale dependent LECs. $n$--loop graphs contribute 
to polynomials $P_m$ with index $m\leq n$, 
with a term proportional to $s^{(n-1)}$. Up to two loops the leading
contributions are given by
\begin{align}
P_0&=0,&P_1&=-6,&P_2&=6N.
\end{align}
In the following, we make use of the fact that $H(s)$ is analytic in the
complex $s$-plane, cut along the real positive axis. 
Unitarity of the S-matrix determines the discontinuity of $H$ 
across this cut,
\bea\label{eq:discH}
H(s+i\ep)-H(s-i\ep) &=& i\sum_n(2\pi)^4\delta^{(4)}(P_n-p)|\gla 0|S^0(0)|n\gra|^2,\nn
&=& \frac{3i}{16 \pi}|F(s)|^2+\cdots,
\eea
where $F(s)$ is the scalar form factor, 
\beq
\langle 0|S^0(0)| \pi^i(p) \pi^k(p')\rangle
=\delta^{ik}F(s)\,\,,\,\, s=(p+p')^2.
\eeq
 The
ellipsis in (\ref{eq:discH}) denotes intermediate states with more than
 two pions. The main point is the observation that knowledge of the
 discontinuity allows one to reconstruct the leading logarithmic term easily,
by constructing a function with the given discontinuity. 
 This is trivial for a function
 with the structure displayed in (\ref{eq:Hpol}).
Assuming that only two pion intermediate states contribute to the leading
chiral logarithm\footnote{This assumption will be justified below.}, the
relation (\ref{eq:discH}) allows us to calculate the $N$--loop 
leading logarithm 
of the scalar two--point
function, once the $N-1$ loop leading logarithm of the scalar form factor is
given. As the leading logarithm to two loops of the scalar form factor is
known \cite{GasserMeissner,BGT}, the leading contribution of the polynomial $P_3$ can be
calculated along the lines just mentioned,
\beq
P_3 = -\frac{61}{9}N^2.
\eeq
This result is in agreement with the relation (\ref{eq:A3}) obtained from
renormalization group arguments, and justifies the neglection of four pion
intermediate states in the unitarity relation (\ref{eq:discH}). In appendix
\ref{app:is}, a more general argument is given.

This method allows one to calculate quite easily
also higher--order leading logarithms. Indeed, as just seen, the scalar form factor
$F(s)$ determines the discontinuity of $H(s)$. Unitarity may again be used to
determine chiral logarithms in $F(s)$, and hence in $H(s)$. Let us illustrate the
procedure by determining the leading term in the polynomial $P_4$ in
(\ref{eq:Hpol}), which amounts to the determination of the LL at 
four--loop order. 
The scalar form factor is analytic in the complex $s$-plane, cut along the
 positive real axis. Its chiral expansion can be arranged in the same way as
 for the correlator in Eq.~(\ref{eq:Hpol}),
\beq
F(s) = 2B \left\{T_0+T_1 L + T_2 L^2 + \cdots \right\},
\eeq where the coefficients $T_i$ are again dimensionless polynomials in $N$ and in the
LECs. The leading contributions to the tree--level, one-- and two--loop result are given by
\begin{align}
T_0&=1,&T_1&=-N,&T_2&=\frac{43}{36}N^2.
\end{align}
The discontinuity across the cut of $F(s)$ is given by
\beq\label{eq:discF}
F(s+i\ep)-F(s-i\ep) = 2i\, t^0_{\,0}(s) F^*(s)+\ldots.
\eeq
The quantity $t^0_{\,0}$ is the isospin $I=0$ S-wave of the
$\pi\pi$-scattering amplitude, in the notation of Ref.~\cite{GL1983}.
The ellipsis in (\ref{eq:discF}) denotes contributions from intermediate states
with more than two pions. Again discarding these for the moment,
the relation (\ref{eq:discF}) shows that knowledge of the LLs of both, 
the scalar form factor and
$t^0_{\,0}$ at $N$ loops, determines the LL of the scalar form
factor at $N+1$ loops, which then determines the LL in $H(s)$ to $N+2$ loops.

Here, we use the fact that $t^0_{\,0}$ is known up to 
two loops \cite{piscattering}. Together with the
two--loop leading logarithm of $F(s)$ \cite{GasserMeissner,BGT}, one finally finds
the three--loop result for $F(s)$ and the four--loop result for $H(s)$, respectively,
\begin{align}
T_3 &= -\frac{143}{108}N^3 + O(s^4), &P_4 &=
\frac{68}{9}N^3+O(s^4).
\end{align} Again, we explicitly checked that $T_3$ is correct by using the renormalization group
equation of \cite{colangelobuechler} as presented in Sect.~\ref{sect:oneloop}.

\section{Leading logarithms to five--loop order}
The arguments in the previous section show that the correlator $H$, the scalar
form factor $F$ and the isospin zero S-wave determine a closed
system as far as LLs are concerned. The chain is
\beq\label{eq:chainunit}
t_{\, 0}^0(s) \stackrel{\mathrm{disc}}{\longrightarrow} F(s)\stackrel{\mathrm{disc}}{\longrightarrow} H(s)\,,
\eeq
 where each step increases the knowledge of the LL by one--loop. The
 analytic structure of $t^0_{\,0}$ differs from the one of $H$ and $F$. 
While the 
correlator $H$ and the scalar form factor $F$ are holomorphic 
 in the complex $s$-plane cut along the positive real axis, this is not the
case for $t_{\, 0}^0$. Indeed, this partial wave is holomorphic in the complex
$s$-plane cut along the real axis in the intervals $(-\infty,0)$ and 
$(4M_\pi^2,\infty)$ for non vanishing quark masses. 
In the chiral limit considered here, the branch points at 
$s=0$ and $s=4M_\pi^2$ coalesce: the partial wave contains, near $s=0$,
singularities of the type $L_- = \ln{-\tfrac{s}{\mu^2}}$ as well as of the
type $L_+ = \ln{\frac{s}{\mu^2}}$. The latter are generated by the left hand
cut. \newline
As a result of this structure, unitarity alone 
does not provide sufficient information to determine the LL in $t_{\, 0}^0$.
 We illustrate the problem at the two--loop level:
The structure of LL-terms of $t^0_{\, 0}$ reads at this order
\beq
t^0_{\,0}(s) = \frac{s^3}{F^6}\{a  L_-^2+bL_+ L_- + c L_+^2  \}+ \cdots.
\eeq
To calculate the discontinuity of $F(s)$ according to Eq.~(\ref{eq:discF}), only the sum
of the coefficients of all leading logarithms is required. It
is determined by the difference between the imaginary part for $s = -|s|+i\ep$,
\mbox{$\mbox{Im}^- t^0_{\,0} = \tfrac{s^3}{F^6} i \pi  (b+2c)L_+$} and the
imaginary part for $s = |s|+i\ep$, $\mbox{Im}^+ t^0_{\,0} = -\tfrac{s^3}{F^6}i \pi (2a+b) L_+$.\newline
Using unitarity restricted to two pion intermediate states for $t^0_{\, 0}$ 
\beq
t^0_{\, 0}(s+i\ep)-t^0_{\, 0}(s-i\ep) = 2i \left| t^0_{\, 0}(s)\right|^2 +\cdots,
\eeq
only determines the imaginary part for $s > 0$. 
However, there exists a set of integral
equations for the partial wave amplitudes of elastic $\pi \pi$ scattering, the
Roy equations \cite{Roy}. These equations provide the necessary information to
determine the left hand singularity. \newline
This allows us to calculate the leading part of the polynomial
$P_5$. Let us switch on the quark masses for the moment.
The Roy equations are of the form 
\beq
t^I_{\ell}(s) = k^I_{\ell}(s)+\sum_{I'=0}^2 \sum_{\ell'=0}^{\infty} \int_{4
  \Mpi^2}^{\infty} ds' K^{II'}_{\ell \ell'}(s,s')\mbox{Im}\, t^{I'}_{\ell'}(s').
\eeq where $k^I_{\ell}(s)$ is a subtraction polynomial and $K^{II'}_{\ell
  \ell'}(s,s')$ are known functions. To determine the left hand cut, knowledge
of the imaginary part of $t^0_{\, 0}(s)$ for negative values of $s$ is
sufficient. One can therefore drop the subtraction polynomial $k^0_{\, 0}(s)$
and one only needs the imaginary parts of the integration kernels as 
given in appendix \ref{app:roy}, which only differ from zero in a
finite interval of $s'$. As the integration over $s'$ only
runs over positive values of $s'$, unitarity provides the imaginary parts of
the partial waves in the integrand,
\bea\label{eq:roy2}
\mbox{Im}^- t^0_{0}(s) = \sum_{I'=0}^{2}\sum_{\ell'=0}^{\infty}\int_{4 \Mpi^2}^{4\Mpi^2-s}
ds' \mbox{Im}\,K^{0I'}_{0 \ell'}(s,s') \mbox{Im}^+t^{I'}_{\ell'}(s').
\eea
As we are only interested in terms which generate leading logarithms we focus
on parts of $\mbox{Im}t^I_{\,\ell}$ which are proportional to
$L_+^2$.\\
We extend the function $\mbox{Im}\,K^{0I'}_{0 \ell'}(s,s')$ analytically to
the whole complex $s$-plane and choose $s = -4\Mpi^2-\delta+i\ep$ with
$\delta > 0$ and $\ep > 0$. This allows to perform the chiral limit and still
remain in a region where the real part of $s$ is negative. In the infinite
sum, only terms with $\ell' \le 2$ contribute to the leading
logarithm. 
Following once again the established path and using the unitarity relations several times, one
obtains the four--loop LL of the scalar form factor $T_4$ and
the five--loop LL of the scalar two--point function $P_5$,
\begin{align}
T_4 &= \frac{15283}{9720}N^4+O(s^5), &P_5 &=
-\frac{140347}{16200} N^4 +O(s^5).
\end{align}
We checked the sum of the coefficients of the leading logarithms of $t^0_{\,0}$ to the order
$p^{8}$ with the renormalization group \cite{colangelobuechler}.

We shortly mention a technical complication that arises while using the Roy
equations in the chiral limit. According to Eq.~(\ref{eq:roy2}), the integrand
is proportional the product of the imaginary parts of the Roy-kernels and of
the partial waves. The centrifugal barrier disappears in
the chiral limit, as a result of which  the imaginary parts are all 
of the same order at threshold, independent of the angular momentum involved,
\begin{align}
\mbox{Im}t^{I'}_{\,\ell'}(s')\,\, \sim s'\,^4, \qquad \ell'\geq 2\,.
\end{align}
On the other hand, the imaginary parts of the kernels behave like
\begin{align}
\mbox{Im}K^{II'}_{\ell\ell'}(s,s')&\sim\frac{1}{s}\left(\frac{s}{s'}\right)^{\ell'}
\end{align}
in the chiral limit.
 For $\ell' \geq 5$, the integrand thus becomes infrared 
singular and non integrable.
 For $\ell' = 5$, a divergence of the form $s^4
\ln(\Mpi)$ is generated. Divergences of the form
$s^{\ell'-1}\Mpi^{-2 \ell'}$ also occur -- these are, however, 
at least of order $O(s^5)$ and do not
affect the four--loop leading logarithm $T_4$. The divergences never 
appear in the terms relevant for the calculation of
the leading logarithms and do not, therefore, affect 
their coefficients.
\begin{figure}
\begin{center}
\begin{tabular}{cc}
\includegraphics[height=6.6cm,angle=-90]{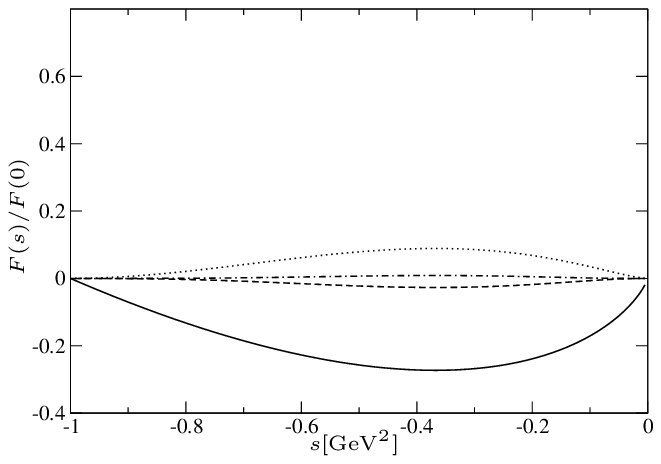}&\includegraphics[height=6.6cm,angle=-90]{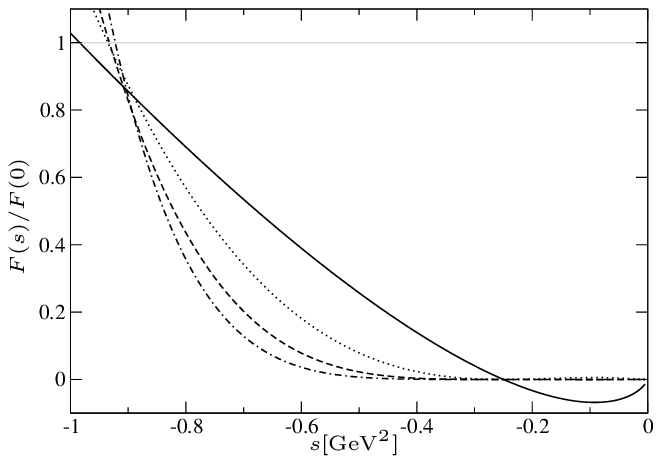}\\
{$\scriptstyle \mu = 1 \GeV$}&{$\scriptstyle  \mu = 500 \MeV$}
\end{tabular}
\caption{Contribution of the LLs to the normalized scalar form factor in a
  region where $F(s)$ is real. The solid, dotted, dashed and dashed-dotted
  lines denote the one--loop, two--loop, three--loop and four--loop LLs,
  respectively. The solid grey line indicates the tree--level result.\label{fig:numerics}}
\end{center}
\end{figure}

To illustrate the size of the LLs, we plot in
Fig.~\ref{fig:numerics} the numerical contributions of the LLs of the scalar
form factor for two choices of the scale $\mu$. One sees that for $s > -0.85 \GeV^2$, the LL
corrections of higher--orders are always suppressed with respect to the lower orders.

\section{Pad\'e approximants}

Pad\'e approximants may be used to estimate higher--order  terms.
The Pad\'e approximant $[M,N]$ for the scalar form factor is 
\beq
[M,N] = \frac{\Xi_M}{\Xi_N} = \frac{F(s)}{F(0)}+O(p^{2(M+N)+2}),
\eeq
where $\Xi_K$ only contains terms up to and including order $p^{2K}$.
In Ref.~\cite{GasserMeissner}, it is shown that the Pad\'e approximant $[0,1]$
does not reproduce the correct factor of the two--loop LL. This shortcoming
persists for the factors of the higher LL. Also the Pad\'e approximants of the two--
and three--loop LLs, $[1,1]$, $[0,2]$, $[2,1]$, $[1,2]$ and $[0,3]$, fail to
reproduce the correct three-- and four--loop LLs, respectively. In
Tab.~\ref{tab:pade}, we compare the two-- and  three--loop Pad\'e approximants
with the exact result. 
\begin{table}[H]
\caption{Pad\'e approximants for the scalar form factor. Displayed are the
  exact and Pad\'e values for the coefficients of the leading logarithms.\label{tab:pade}}
\begin{center}
\renewcommand{\arraystretch}{1.4}
\begin{tabular*}{\textwidth}{c@{\extracolsep{\fill}}ccccccc}
\hline
&$\frac{F(s)}{F(0)}$& $[0,1]$& $[1,1]$ & $[0,2]$ & $[2,1]$& $[1,2]$ & $[0,3]$\\ \hline
$\frac{T_3}{N^3}$ & $-\frac{143}{108}$ & $-1$  & $-\frac{1849}{1296}$ &
$-\frac{25}{18}$ & $-\frac{143}{108}$ & $-\frac{143}{108}$ & $-\frac{143}{108}$\\ 
$\frac{T_4}{N^4}$ & $\frac{15283}{9720}$ & $1$ & $\frac{79507}{46656}$ &
$\frac{2101}{1296}$ & $\frac{20449}{13932}$ & $\frac{1961}{1296}$ &
$\frac{1933}{1296}$\\ \hline
\end{tabular*}
\end{center}
\end{table}

\section{Summary and conclusion}
1. In this letter, we  study the structure of chiral logarithms in 
the two-flavour sector,
 in the chiral limit $m_u=m_d=0$. In particular, we
investigate the chiral expansions of the scaler form factor of the pion, 
and of the correlator of two
isoscalar quark currents. 
We confine our interest to the so called {\it leading}
chiral logarithms, which are defined to be 
the ones accompanied with the least power of external momenta.

2. Standard techniques to calculate these logarithms -- direct evaluation by
calculating Feynman diagrams, or using  renormalization group
techniques -- cannot be applied beyond three--loop accuracy, because the
required labor becomes prohibitive.

3. We point out  that, using unitarity and analyticity, the leading
logarithms for the mentioned correlators can be calculated to any desired
order with a modest calculational cost. We illustrate this claim 
with an evaluation of the leading logarithm for the scalar two--point function to five--loop order.
As far as we are aware, this is the first evaluation of chiral
logarithms at this accuracy. The numerical structure of the 
coefficients appears erratic to us.

4. The proposed technique makes use of Roy-equations for the $\pi\pi$ 
 amplitude near threshold, in the chiral limit. 
While the Roy-kernels become singular in
this limit, we argue (and explicitly check in one case) that 
these singularities do not affect the evaluation of the leading logarithms.

5. We compare our results with  Pad\'e approximants and confirm an
earlier statement \cite{GasserMeissner} concerning  
the use of this technique for  the evaluation of chiral logarithms:
 knowledge of the coefficients up to and including those of order $N$ does
 not allow one to calculate the ones at order $N+1$ and higher.

\section{Acknowledgements}
It is a pleasure to thank G.~Colangelo, C.~Greub, H.~Leutwyler, U.G.~Meissner
and all the members of the institute for informative discussions. We are
indebted to J.~Gasser for useful discussions and a careful reading of the
manuscript. This work was supported in part by the Swiss National Science Foundation,
by RTN, BBW-Contract No. 01.0357, and EC-contract HPRN-CT2002-00311 (EURIDICE).

\begin{appendix}

\section{Roy equations}\label{app:roy}
Using the integral representation for the kernels $K^{I I'}_{\ell
  \ell'}(s, s')$ given in \cite{ACGL} it is straightforward to
calculate their imaginary parts,
\beq
\mbox{Im}\, K^{I I'}_{\ell \ell'}(s,s') = C_{su}^{II'}\frac{4 \ell'+2}{s-4\Mpi^2}
P_{\ell}\bigg(\frac{s+2s'-4 \Mpi^2}{4\Mpi^2-s}\bigg)P_{\ell'}\bigg(\frac{2s+s'-4
  \Mpi^2}{4\Mpi^2-s'}\bigg),
\eeq
where $C_{su}^{II'}$ are the matrix elements of the crossing matrix $C_{su}$ and $P_{n}(x)$ are the
Legendre polynomials. Note that it is crucial to use the projection interval
$[0,1]$ in the integral representation of the kernels as given in
\cite{ACGL}. Otherwise, the Roy equations cease to be valid for negative
values of $s$.\footnote{We are indebted to H.~Leutwyler for discussions on
  this issue.}

\section{Suppression of higher intermediate states}\label{app:is}

In this appendix, we argue that intermediate states with more than two pions
do not contribute to the leading logarithms of the scalar two--point function.
In the sum
\bea
H(s+i\ep)-H(s-i\ep) &=& i\sum_n(2\pi)^4\delta^{(4)}(P_n-p)|\gla 0|S^0(0)|n
\gra|^2
\eea
appear the matrix elements of the $2 k$ pion intermediate states
\beq
Y_{2k \pi} = \gla 0 | S^0(0) |2 k \pi \gra.
\eeq
The tree--level results of these matrix elements start with $F^{-2(k-1)}$. Because the
integrand of the phase space integral is the modulus squared of $Y_{2k \pi}$,
the lowest order contribution of the $2k$ pion intermediate states to the
discontinuity of $H(s)$ is therefore of order $F^{-4(k-1)}$ and does not contain
any logarithms. The loop corrections to these matrix elements are suppressed even
stronger in $F^{-2}$.
Take for example $k = 2$. The four particle phase space integration would have to generate a
logarithm squared to contribute to the leading term in the polynomial
$P_3$. This is not possible and one way to see it is the following:
Only diagrams with vertices solely from $\mathcal{L}_2$ contribute to the
leading logarithms. The highest power of the pole term $\ep^{-1}$ is directly related to
the highest power of the logarithm. 
Since the phase space integration in an infrared save theory does not produce a
divergence, a phase space integration does not increase the power
of the pole term and thus neither the power of a logarithm.
In the case of the scalar form factor, the argument is the same. The higher
intermediate states are also suppressed in $F^{-2}$.

\end{appendix}

\bibliographystyle{unsrt}

\end{document}